\documentclass[manuscript, screen]{acmart}

\AtBeginDocument{%
  \providecommand\BibTeX{{%
    \normalfont B\kern-0.5em{\scshape i\kern-0.25em b}\kern-0.8em\TeX}}}

\setcopyright{iw3c2w3g}
\copyrightyear{2022}
\acmYear{2022}
\usepackage{dirtytalk}
\usepackage[utf8]{inputenc}  % translates  standard and other input encodings into a ‘LATEX \usepackage{float} % Improves the interface for defining floating objects such as figures and tables.
\usepackage{subcaption} % captions for subfigures and the like

\usepackage{url}
\makeatletter
\g@addto@macro{\UrlBreaks}{\UrlOrds}
\makeatother

\usepackage{multirow} %tabular cells spanning multiple rows
\usepackage{color} %colour management
\usepackage{enumitem} %better  enumerate, itemize and description

\newcommand{\m}{\textit{M=}}

\newcommand{\sd}{\textit{SD=}}

\newcommand{\N}{\textit{N=}}

\def\plaintitle{Exploring the Social Context of Collaborative Driving}

\def\plainkeywords{Interface design; collaborative driving; study.}

\begin{document}

%[shorttile]
\title{\plaintitle}

\author{Mark Colley}
\email{mark.colley@uni-ulm.de}
\orcid{0000-0001-5207-5029}
\affiliation{%
  \institution{Institute of Media Informatics, Ulm University}
  \city{Ulm}
  \country{Germany}
}

\author{Sebastian Pickl}
\email{sebastian.pickl@uni-ulm.de}
\affiliation{%
  \institution{Institute of Media Informatics, Ulm University}
  \city{Ulm}
  \country{Germany}
}

\author{Frank Uhlig}
\email{frank.uhlig@elektrobit.com}
\affiliation{%
  \institution{Elektrobit Automotive GmbH}
  \city{Ulm}
  \country{Germany}
}

\author{Enrico Rukzio}
\email{enrico.rukzio@uni-ulm.de}
\orcid{0000-0002-4213-2226}
\affiliation{%
  \institution{Institute of Media Informatics, Ulm University}
  \city{Ulm}
  \country{Germany}
}

\renewcommand{\shortauthors}{Colley et al.}

\begin{abstract}
%Motivation
The automation of the driving task affects both the primary driving task and the automotive user interfaces. The liberation of user interface space and cognitive load on the driver allows for new ways to think about driving. 
%Problem
Related work showed that activities such as sleeping, watching TV, or working will become more prevalent in the future. However, social aspects according to Maslow's hierarchy of needs have not yet been accounted for. 
%Solution
We provide insights of a focus group with \N{5} experts in automotive user experience revealing current practices such as social need fulfillment on journeys and sharing practices via messengers and a user study with \N{12} participants of a first prototype supporting these needs in various automation levels showing good usability and high potential to improve user experience.

\end{abstract}



% Author Keywords

\maketitle

\section{Introduction}\label{sec:introduction}
%Motivation
Advanced Driver Assistance Systems and the automation of the driving task~\cite{taxonomy2014definitions} change the interaction with vehicles. Driving task-specific data will likely lose importance to an extent where currently necessary information is no longer needed~\cite{kun2016shifting}. Changing information needs and activities are expected~\cite{pfleging2016investigating, yang2018will}. According to Maslow's hierarchy of needs~\cite{mcleod2007maslow}, when \textit{safety needs} are fulfilled, \textit{love needs} can be considered.\\
%Problem
Currently, user interfaces are designed for a driver fulfilling the driving task~\cite{kern2009design}. While some work explored the driver --- passenger interaction with specialized applications~\cite{perterer2015co, perterer2013come}, interaction across vehicles~\cite{perterer2019driving} was only most recently explored. However, the \textit{love needs}, or \textit{need for social belonging}~\cite{hagerty1992sense}, were not addressed. We expect a rise in addressing these needs with further automation of vehicles. 
While it is not yet clear whether a gradual approach to automation such as done by Tesla or an abrupt change to fully autonomous vehicles will happen, we expect the \textit{love needs} to increase in both cases.
%Solution
Therefore, we implemented an Android application that supports relevant needs such as communication and reassuring the other's position.

In an expert focus group (\N{5}) in the automotive industry, the current social context of driving and needs for future rides were elicited. The prototypical application was then evaluated with \N{12} participants both from the industry as well as from a general population.

\textit{Contribution Statement:} We present the findings of an expert focus group (\N{5}) in the automotive industry and a prototype for the support of the \textit{love needs} applicable for manual and automated driving. The findings of a first exploratory experiment of the application indicate good usability and that the sense of belongingness could be increased.

%\section{Related Work}
%We review approaches and technologies for collaborative driving and expectations for AVs.

\section{Collaborative Driving}
Research into collaborative driving and the social aspects thereof was mostly conducted within one car using an ethnographic approach~\cite{brown2012normal, laurier2008driving, cycil2013eyes, forlizzi2010should, perterer2013come}. Some research specifically focused on navigational tasks~\cite{forlizzi2010should, sheller2000city, trosterer2015four, perterer2015co}. The passenger was found capable of giving relevant additional information~\cite{forlizzi2010should}. This is used in the \textit{shared gaze} approach~\cite{trosterer2015four}, in which the gaze of the passenger is visualized for the driver. It is enhanced in the \textit{Co-Navigator} system~\cite{perterer2015co}. This system aids the passenger in supporting the driver by providing additional information such as landmarks and hazard warnings. Perterer et al.~\cite{perterer2019driving} also explored \say{diving together across vehicles}. In their experiment, participants were positioned in different vehicles and were given four tasks. %: (1) Stop because the trunk is open, (2) change route due to closed highway, (3) stop at a gas station, and (4) stop at a fast-food restaurant.
They investigated communication, context factors, and coping strategies. Direct communication (face-to-face) was employed at the start. Later, the smartphone was used. The position of the front car was used as non-verbal communication. Information conveyed was route instructions, localization, meeting places, and coping strategies such as route selections or defining meeting points~\cite{perterer2019driving}. Relevant social and context factors were \textit{driving habit}, \textit{patience}, \textit{predictability of behavior}, and \textit{distraction through technology}. No specialized applications or technology were used.

\subsection{Expectations on Autonomous Driving}
Non-driving related tasks in autonomous vehicles (AVs) are expected to become more prominent~\cite{pfleging2016investigating, yang2018will}. Expected activities include listening to music, talking, texting, reading, sleeping, smoking, learning, watching movies, or playing video games~\cite{pfleging2016investigating, yang2018will}. Requirements for the interior of AVs: rotatable seats, fully flat seats, retractable steering wheel and more space, table for food and laptop, and enhanced lighting options (e.g., dimmable)~\cite{yang2018will}. Frison et al.~\cite{frison2019you} evaluated the impact of scenario in an AV on psychological needs. 
Frison et al.~\cite{frison2019you} showed that psychological needs vary with the scenario: trust issues are more prevalent in complex scenarios.

\section{Focus Group}
%(see~\autoref{table:participants_focus_group})
We conducted a focus group with \N{5} employees (3 female, 2 male) of \textit{Elektrobit}~\cite{elektrobit} to (1) discuss the social context of driving, (2) to explore use cases, and (3) to evaluate a prototypical implementation. Participants were on average \m{28.60} (\sd{2.97}) years old and had \m{3.20} (\sd{1.64}) years of experience working in automotive user experience.
%, a company working on automotive software

\subsection{Procedure}
The focus group consisted of three phases: \textit{introduction}, \textit{discussion}, and an \textit{evaluation} of a prototype and lasted about 1.5h. During the whole session, audio was recorded.
%The focus group started with an \textit{introduction} of the participants and the research field of the organizer. 
Approaches to navigation such as Waze~\cite{waze}, Wolfpack~\cite{wolfpack}, and Google Maps~\cite{googleMaps} were presented. The evolving automation of the driving task~\cite{sae2014taxonomy}, the resulting reduction of needed space in user interfaces (UI), and freed cognitive resources were highlighted. 
The need for \textit{social belongingness} was justified with Maslow's hierarchy~\cite{mcleod2007maslow}.
Participants were then instructed to propose ideas on fulfilling this need with current driving and future AVs.
Afterward, a video and screenshots of a prototype were shown for feedback. % on design and implementation.
%Results are described in section~\ref{section_results_workshop_one}.
%The focus group lasted about 1.5h. 
No special automation level~\cite{sae2014taxonomy} for the application was given. %, 20min for the \textit{introduction}, 40min for the \textit{discussion} and 20min for the \textit{evaluation}.

\subsection{Analysis}
Discussions of the focus group were analyzed using the method proposed by Mayring~\cite{mayring2015qualitative}: The second author lead the group discussion. Afterward, the first and second authors listened to the recordings independently. They then discussed relevant topics with their associated anchor quotes. Emerging themes were then discussed. %We report these topics in the following.

\subsection{Results}
\textit{Location Sharing:}
All participants agreed \textit{location sharing} to be the most important feature in social navigation. [P1] stated this to be \say{best in real time} to provide information regarding location, directions, and activities.
%be able to answer these questions: \say{where's the other person, where's the other person going, what's the other person doing?} [P4]. 
This seems necessary to keep the group together and provide a sense of not being able to be separated from the group. % (\say{As a driver I would like to see at any time where the individual members of the group are in case someone gets lost} [P4]). It was even said that when traveling in a group, it is more important to stick together than where you are going. % (\say{it doesn't matter as much if you get there as long as you get there together} [P4]).
This should be integrated with classic navigation. % (\say{classic navigation plus the position of the others} [P2]). % The prototype seemed to match the expectations of the group in a way that there was not much discussion about the actual implementation besides some comments about the visual implementation as described under implementation.

\textit{Communication:}
%A major point of discussion was the topic of communication, the exact implementation, and its implications. In the first round of finding ideas, suggestions were a chat functionality, enabling communication between cars (\say{for example, I would like a chat function [...] to communicate with the occupants of the other vehicle [...]} [P1], \say{or also video telephony} [P1], \say{the opportunity to talk to them somehow} [P2], \say{What is important to me: communication} [P5]).
Communication between vehicles was seen as a crucial part.
Suggested methods were text messages, voice messages, video calls, walky-talky like functionality, a permanently open channel for audio communication, or a push-to-mute approach. % (\say{SMS [...] voice messages or permanently open audio channel as in telephony or a walky-talky function where I simply press a button and then the audio channel is open} [P5], \say{Or even video telephony} [P1]).
%While the group quickly focused on audio communication, 
The modalities of audio communication remained the center of discussion. One concern was that a permanent audio connection between cars could lead to information overflow %(\say{permanently open could become too much} [P5]) 
as there are conversations about topics only relevant for passengers of one car. 
Missing awareness for the others` driving situation which could lead to safety risks was also mentioned. %(\say{I think the problem of 'I get when the other one is in an important driving situation' [...] works well in a 1 to 1 relationship but if I imagine there is [...] more than one car, it becomes impossible at some point} [P1]). 
There were also concerns about the safety of using voice messages as they require explicit interaction to listen or record them. % (\say{Voice messages [...] on the one hand I think they are a little distraction, but on the other hand also [...] that it distracts} [P4], \say{you have to explicitly notice that there is a voice message, explicitly press a button that you can hear it, explicitly press a button when you want to record something, maybe that's easier with something open, you just have to talk and listen} [P4] ), but not both distractions were perceived equally bad. 
Explicit interactions only distract the driver if wanted while implicit interaction is uncontrollable (\say{%but possibly these interactions are exactly when the driving situation allows it. 
With open communication, I'm probably less distracted, but exactly where I shouldn't be} [P5]). Therefore, a mute button %(\say{I can also imagine that you could have a prominent button somewhere to pause the communication} [P3]) 
or a driver, driving, or, in higher automation levels, vehicle status indicator %(\say{[...] like Skype} [P2]) 
which could rely on navigation data, was proposed. %as critical situations like upcoming turns could be predicted from maps. 
%The state of the car in higher automation levels could also be communicated. 
\say{Mostly, it [the app] knows when a difficult driving situation is upcoming [...] %for example turning, traffic lights or parking, it would also be interesting if we drive autonomously [...], 
no matter in which [automation] level, that you can communicate this} [P5].

Whether audio messages should be explicit, %designed as an explicit mean of communication,
meaning a trigger is needed or whether an open channel serves the purpose best was a discussion point. Further research seems necessary on the usage of such a tool for brief informational messages or longer conversations. %(\say{for some people it is strange to talk a lot, it would be a short thing, much more informative, some will use it very differently and talk a lot} [P4]).
Due to the general characteristics of a communication tool, it was perceived as the solution for a lot of other imagined requirements % the participants had when they were asked for ideas. 
including the possibility to share route changes, breaks, and points of interest. % (POI) (\say{the possibility to share intermediate stops with others [P2], quick appointment and determination of common breaks [P1], this POI thing you can easily communicate visually} [P4]).

\textit{Media Sharing and Entertainment:}
%The third feature idea was sharing media and entertainment. 
An example for shared media consumption to intentionally create the illusion of togetherness was given by [P5]: \say{if I go on holiday with my girlfriend, we choose a film together in the plane, [...] %if we have this infotainment system,
then we usually sit there, count down, 1-2-3 and then click simultaneously on play [...], %to watch the film together, 
that's somehow something different than if I were 10 seconds delayed}. The wish for a shared media context also extended to music currently played or sharing experiences. %(\say{Media sharing... Music, titles that are currently playing} [P1], \say{as a driver I would like to share experiences with the drivers of other vehicles} [P4]). 
Playing driving-related games like battles about speed or economy also merged into the context of shared media and entertainment (%\say{in my head I also had [...] games} [P4], 
\say{I would like to have competition [...] who has the maximum speed, who has the most speed overruns, who was the most fuel efficient} [P1]).

\textit{Two Aspects of Safety:}
%Safety was a major concern in feature design and implementation. 
%Participants worried about distractions threatening safety.
For entertainment and competitive games, participants question responsibility for following the traffic rules when everyone tries to break some kind of score (\say{who follows traffic rules? [...]} [P5]). % who has the maximum speed, who has the most speeding violations, who was the fuel saver} [P5]). This perspective on safety was also predominant in the discussion on communication channel design.
Safety was also interpreted as safety from losing connection to other cars in the group, mainly visually (\say{Need for safety: where are the others?} [P4]). This could go so far that the other aspect of safety, physical integrity, is put at risk by ignoring traffic rules to stay in visual contact (\say{you drive easier over a red or yellow traffic light because you want to stay behind the others} [P4]). %[P5] stated that \say{this gives the driver safety when I know: where to go, where are the others?}

\textit{Together vs. Individually:}
An important aspect of social navigation were boundaries: what should be shared and what should remain in individual contexts. Sharing as much information as possible was not approved. It was noted that people choose a car based on the music played. % (\say{we drive different cars so we can hear different things} [P4], \say{with music, one argues rather more} [P4]), while others also enjoy listening to music together (\say{I bought a Y-connector for my mobile phone so that you can plug in 2 headphones to listen to music together} [P5]). 
In general, the \textit{common} experience is important to the participants (\say{I care about what we have in common} [P5]). Situations in which splitting into separate cars creates group dynamics like insider information that even stays intact and keep the group divided after reunification (\say{I have the experience [...] each car is its group, has its insiders [...] the group is so split afterwards} [P3]). That is why the wish for features that break the borders between cars and transport the feeling of not being separated is so dominant (\say{border of the two cars breaks up and a feeling of not being separated builds up} [P5]), for example by communicating currently playing music or videos. % (\say{communicate the same music or videos} [P5]). 
This already happens without the existence of supporting technology by sharing media links via messengers (\say{with my friends [...] hey, we hear that now and then we send a link} [P2]).
%just by using existing features like sharing links to media via messengers (\say{with my friends [...] hey, we hear that now and then we send you a link} [P2]).

\section{Experiment}
We conducted an experiment to evaluate the basic concept of collaborative driving.
A microphone was used for voice recording. 
%Screenshots are depicted in~\autoref{fig:app_screenshot} and~\autoref{fig:app_screenshot_2}.

\subsection{Apparatus}
\begin{figure}[ht!]
    \centering
     \begin{subfigure}[b]{0.33\textwidth}
        \centering
        \includegraphics[width=0.75\textwidth]{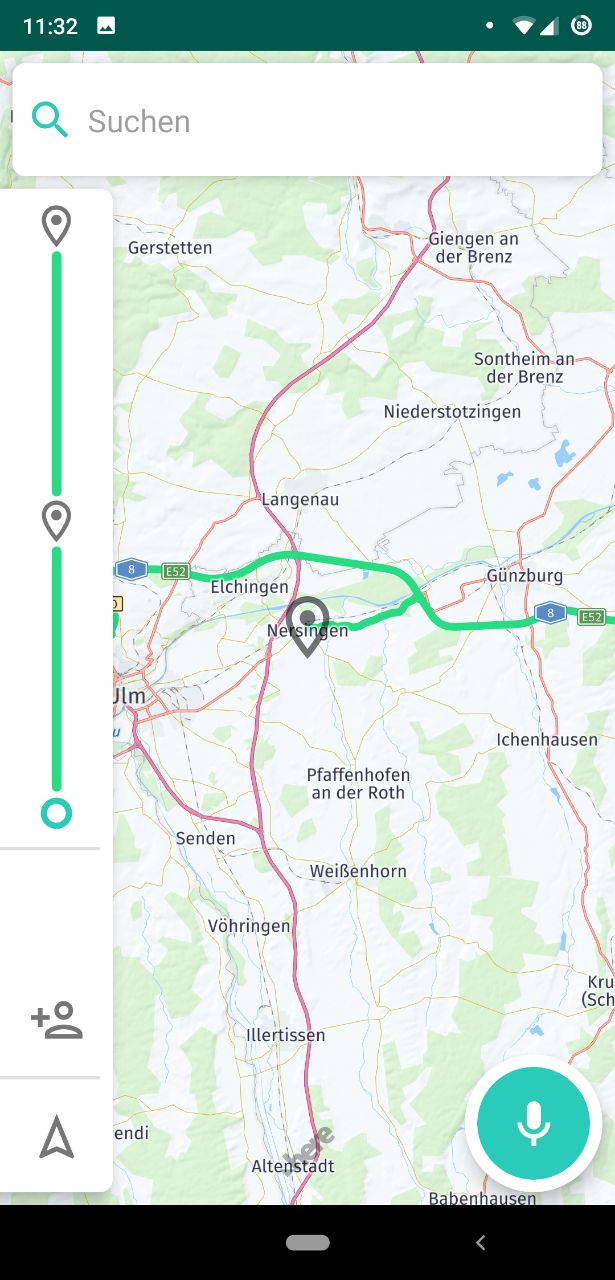}
        \subcaption{Screenshot of the application. No friend invited. Drawer is closed, on the right side is the button for communication.}
        \label{fig:app_screenshot_2}
    \end{subfigure}
        \begin{subfigure}[b]{0.33\textwidth}
        \centering
        \includegraphics[width=0.75\textwidth]{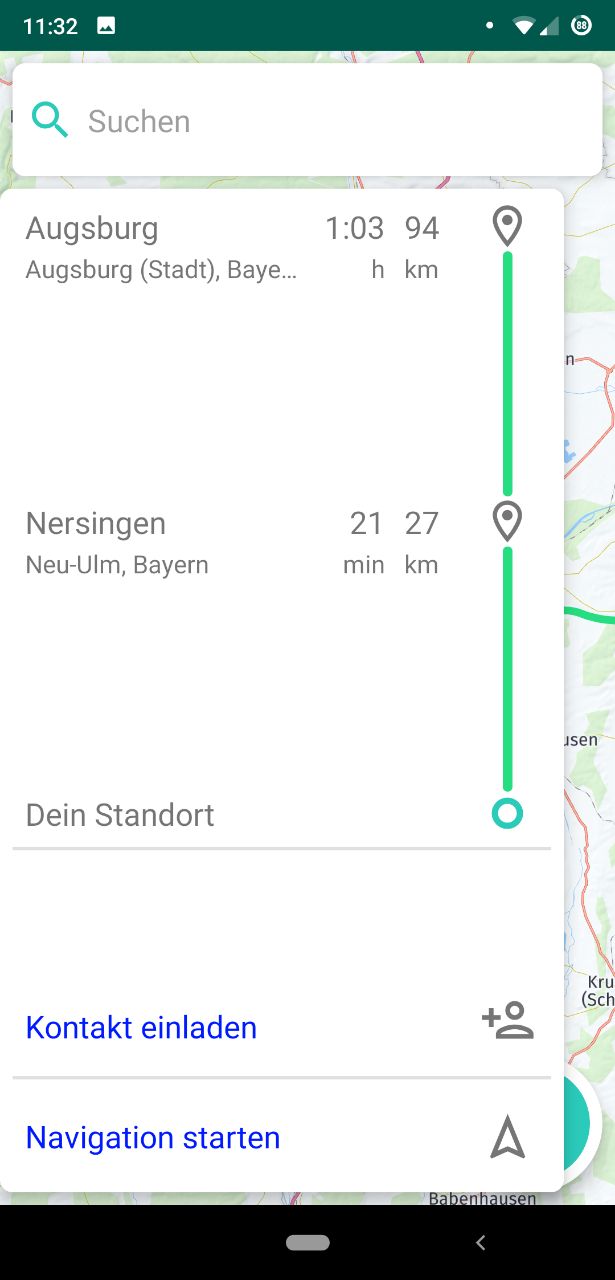}
        \subcaption{Screenshot of the application. No friend invited. Drawer is opened.\newline}
        \label{fig:app_screenshot}
    \end{subfigure}
            \begin{subfigure}[b]{0.33\textwidth}
        \centering
        \includegraphics[width=0.75\textwidth]{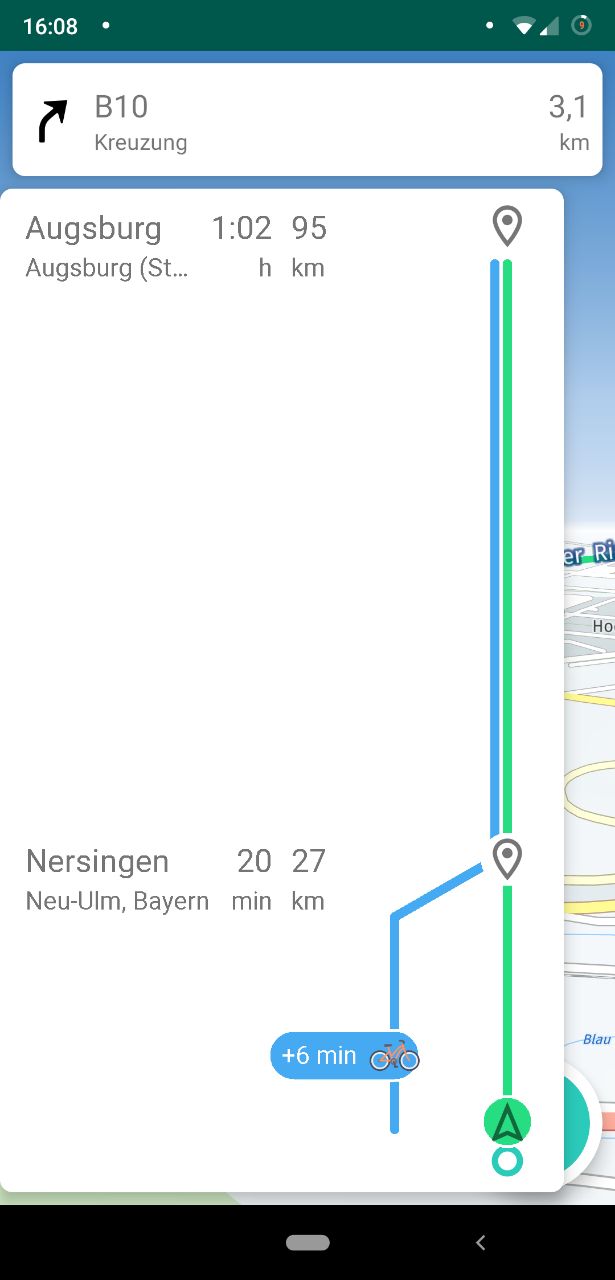}
        \subcaption{Screenshot of the application. One friend invited. Drawer is opened with a merging route.}
        \label{fig:app_screenshot_3}
    \end{subfigure}
    \caption{Screenshots of the application.}
    \label{fig:app_screenshots}
    %\vspace{-5mm}
\end{figure}

\begin{figure}[ht!]
    \centering
        \includegraphics[width=0.5\textwidth]{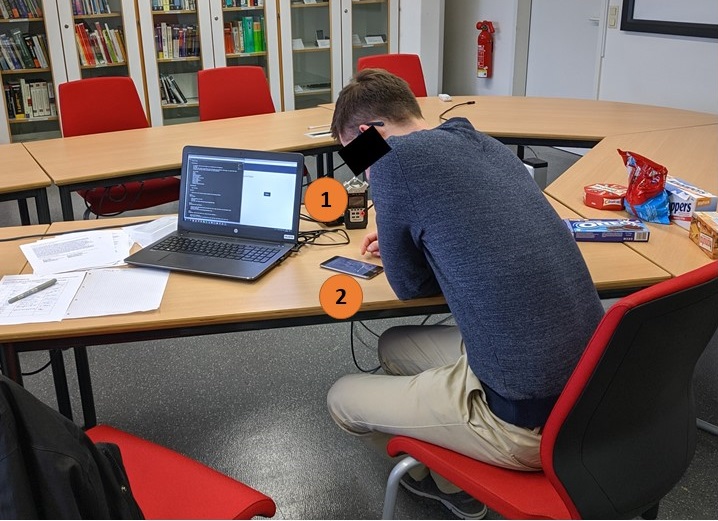}
        \caption{One participant engaged in the app. }
        \label{fig:apparatus}
\end{figure}

We developed a prototype for a collaborative navigation application based on the expert focus group as shown in~\autoref{fig:app_screenshots}. 
This application allows for \textit{inviting friends} in the navigation context, continuous \textit{location sharing}, \textit{collaborative modification of route}, and \textit{walky-talky} like communication. \autoref{fig:app_screenshot_3} shows the train metaphor used to visualize distances and common and separate parts of the route. The advantages of this design are simplicity of the abstract route representation due to the elimination of redundancies while still preserving proportionality.
For the study, participants sat in a meeting room displayed in~\autoref{fig:apparatus}. The application simulated a ride towards a destination.

\paragraph{Participants}
We recruited \N{12} participants (4 female, 8 male) that were on average 26.33 years (\sd{3.39}) old. 6 participants were recruited from Elektrobit (called \textit{Experts}), 6 from the university population (called \textit{laypersons}).

\paragraph{Procedure}
After giving informed consent, participants were introduced to the app and its functionalities: map view, search, add/delete destination, voice messaging, and . Afterward, three tasks were explained:
(1) Set location, invite friend, wait for a friend, add waypoint (Wizard of Oz).
(2) Send a message for a friend to remove waypoint.
(3) Start navigation, locate other participant, describe the relative location.

Finally, a questionnaire including the self-assessment manikin (SAM)~\cite{bradley1994measuring}, the  post-study system usability questionnaire (PSSUQ)~\cite{lewis1992psychometric} (lower is better), and demographic questions were filled out. Open feedback was also requested. On average, a session lasted 18 min. All participants received sweets as compensation.

\section{Results}
We used Mann-Whitney or Welch tests depending on the data's nature~\cite{spss_statistics_assumptions} to compare layperson and expert responses. We used Version 4.1.2 of R with all packages up-to-date as of January 2022. RStudio Version 2021.09.0 was used. For the figures, we used the package \textit{ggstatsplot}~\cite{ggstatsplot} in version 0.9.1.

%\paragraph{Affective State:} %Affective state was measured using the SAM~\cite{bradley1994measuring}. 

\begin{figure}[ht!]
    \centering
     \begin{subfigure}[b]{0.33\textwidth}
        \centering
        \includegraphics[width=\textwidth]{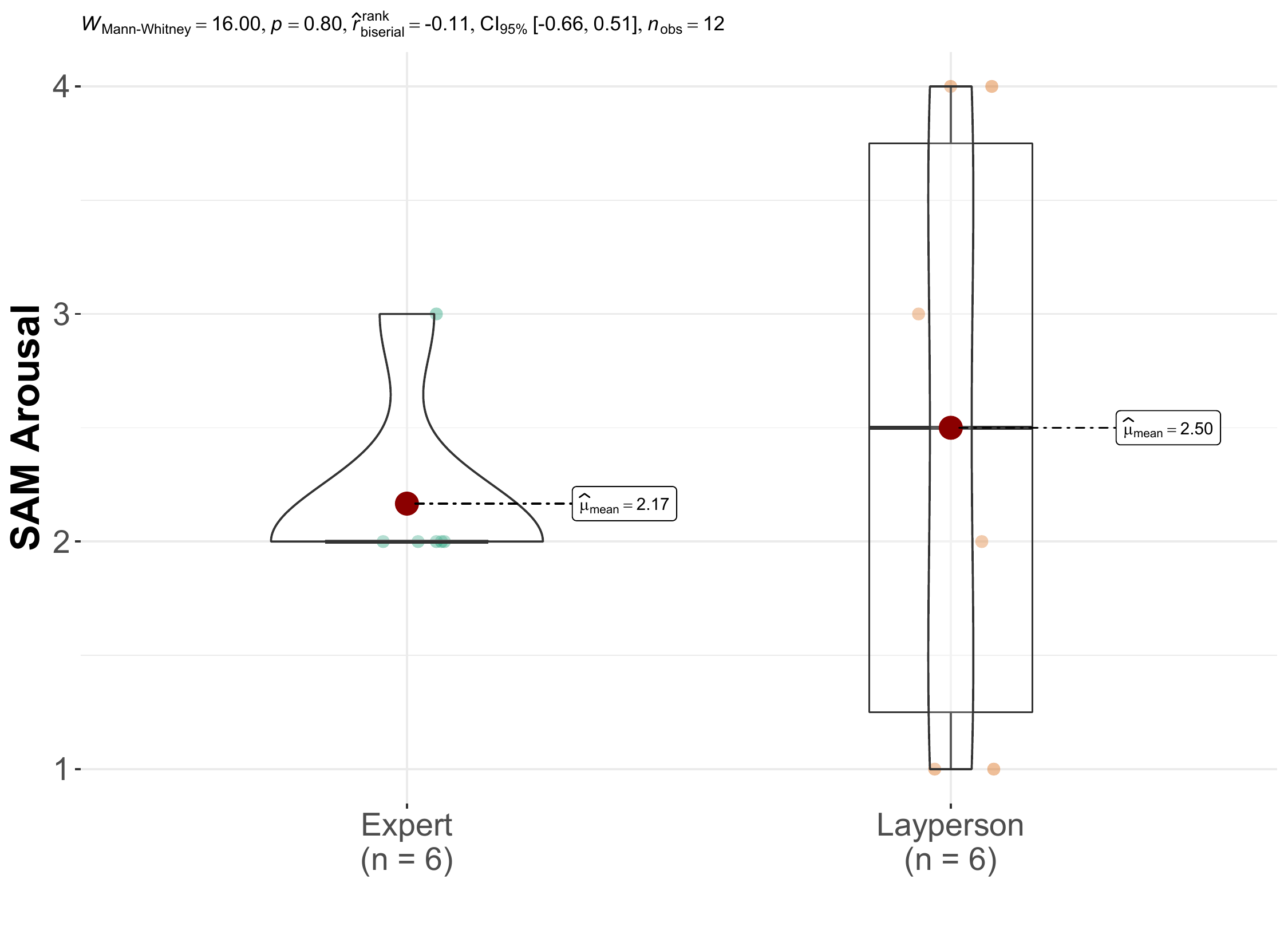}
        \subcaption{SAM Arousal}
        \label{fig:sam_arousal}
    \end{subfigure}
        \begin{subfigure}[b]{0.33\textwidth}
        \centering
        \includegraphics[width=\textwidth]{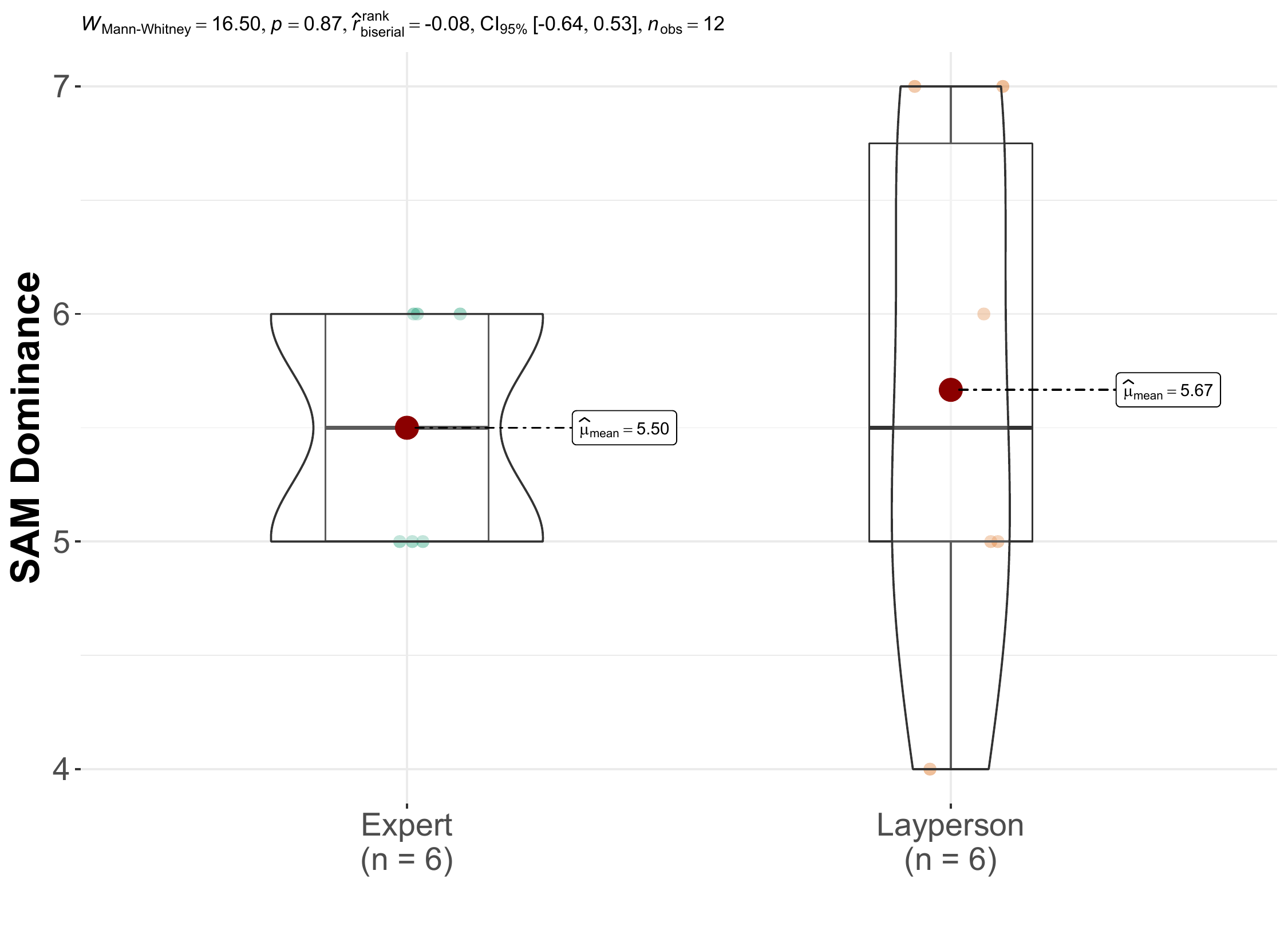}
        \subcaption{SAM Dominance}
        \label{fig:sam_dominance}
    \end{subfigure}
    \begin{subfigure}[b]{0.33\textwidth}
        \centering
        \includegraphics[width=\textwidth]{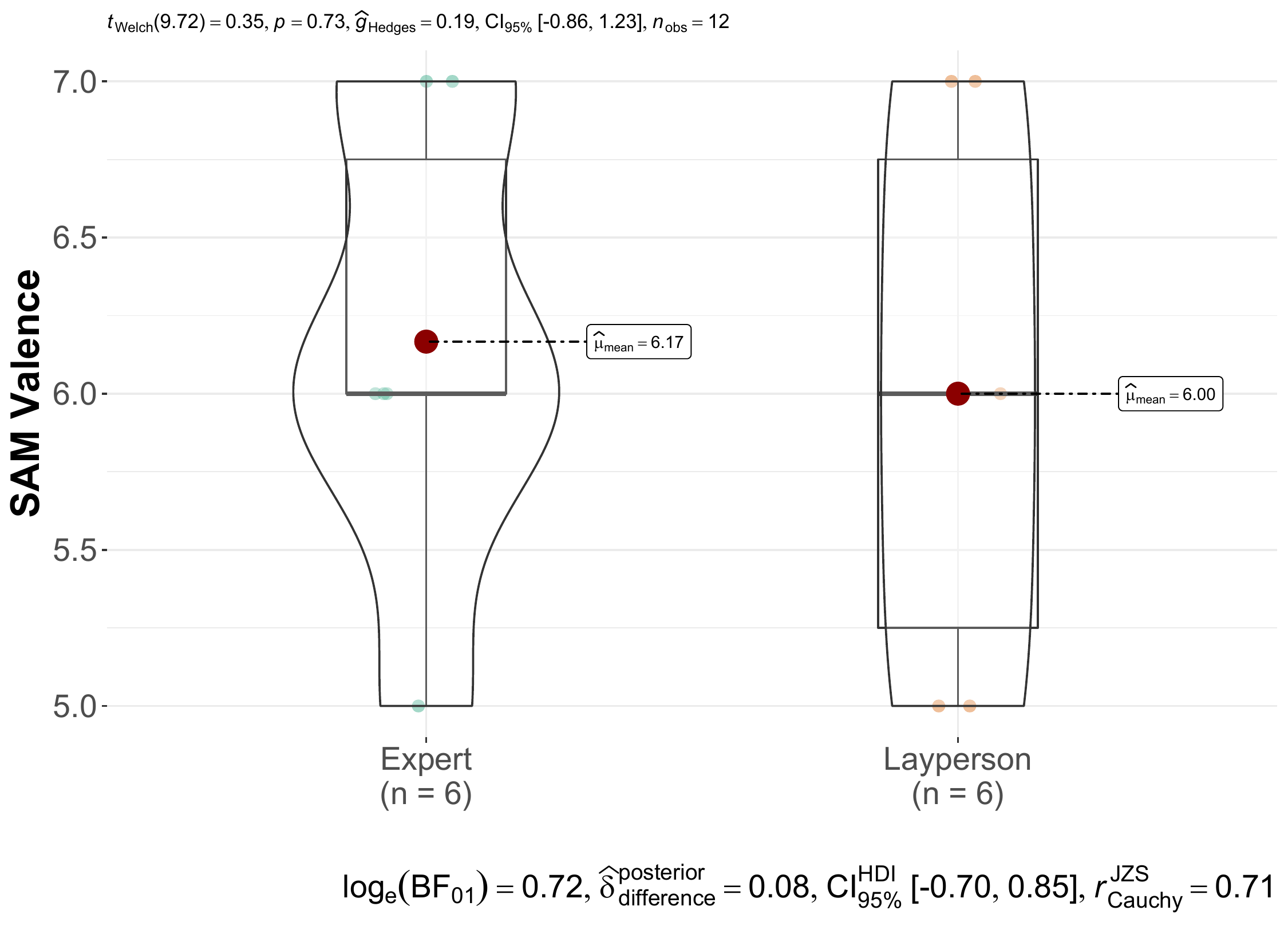}
        \subcaption{SAM Valence}
        \label{fig:sam_valence}
    \end{subfigure}
    \caption{Comparison between layperson and experts for the SAM~\cite{bradley1994measuring}}
    \label{fig:sam_scores}
\end{figure}

\paragraph{Measurements:}
Reported values for Dominance (\m{5.58}, \sd{0.90}), Valence (\m{2.33}, \sd{0.98}), and Arousal (\m{6.08}, \sd{0.79}) indicate high approval. No significant differences were found between experts and layperson (see~\autoref{fig:sam_scores}).

\begin{figure}[ht!]
    \centering
     \begin{subfigure}[b]{0.45\textwidth}
        \centering
        \includegraphics[width=\textwidth]{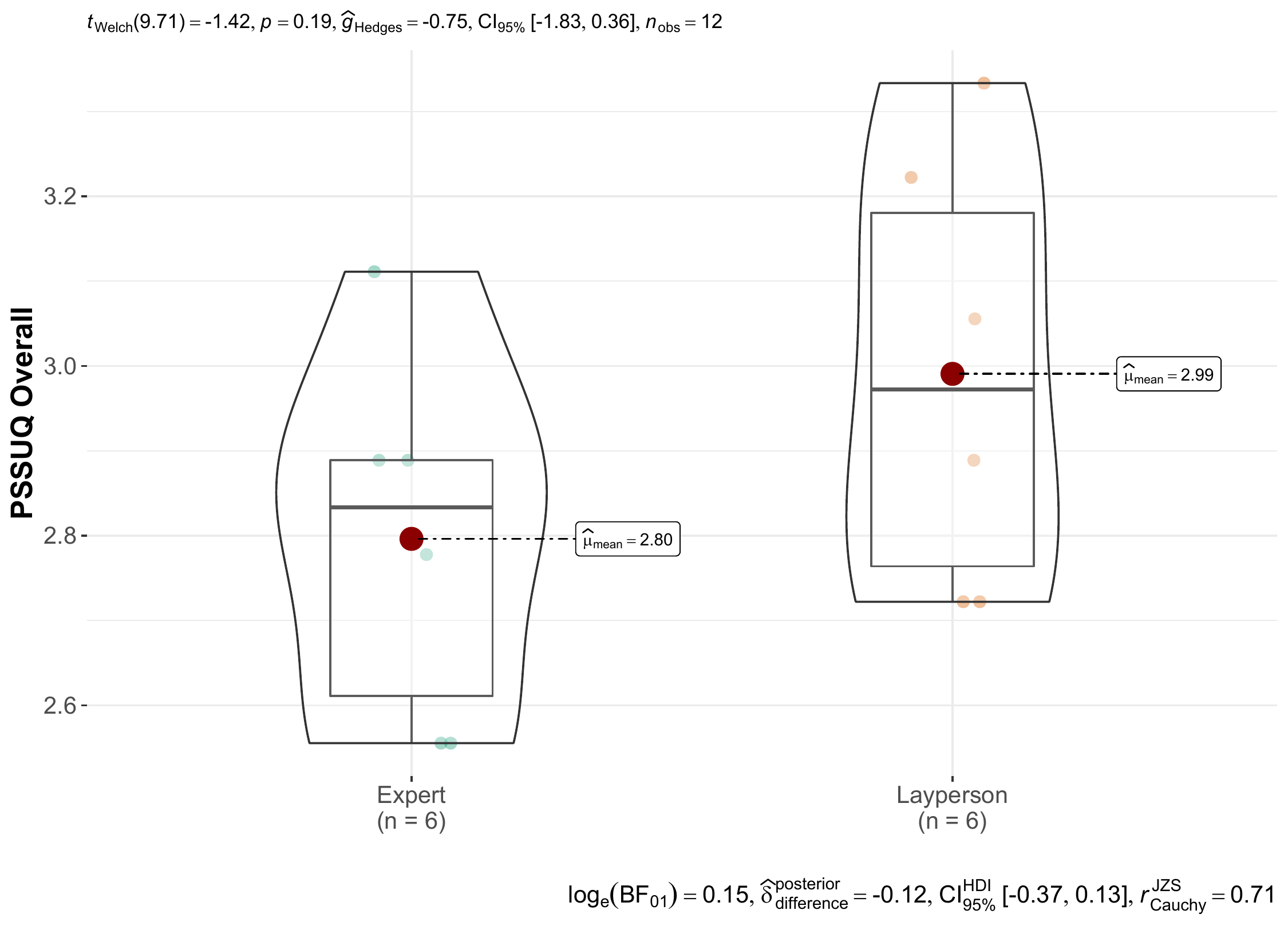}
        \subcaption{PSSUQ Overall}
        \label{fig:pssuq_overall}
    \end{subfigure}
        \begin{subfigure}[b]{0.45\textwidth}
        \centering
        \includegraphics[width=\textwidth]{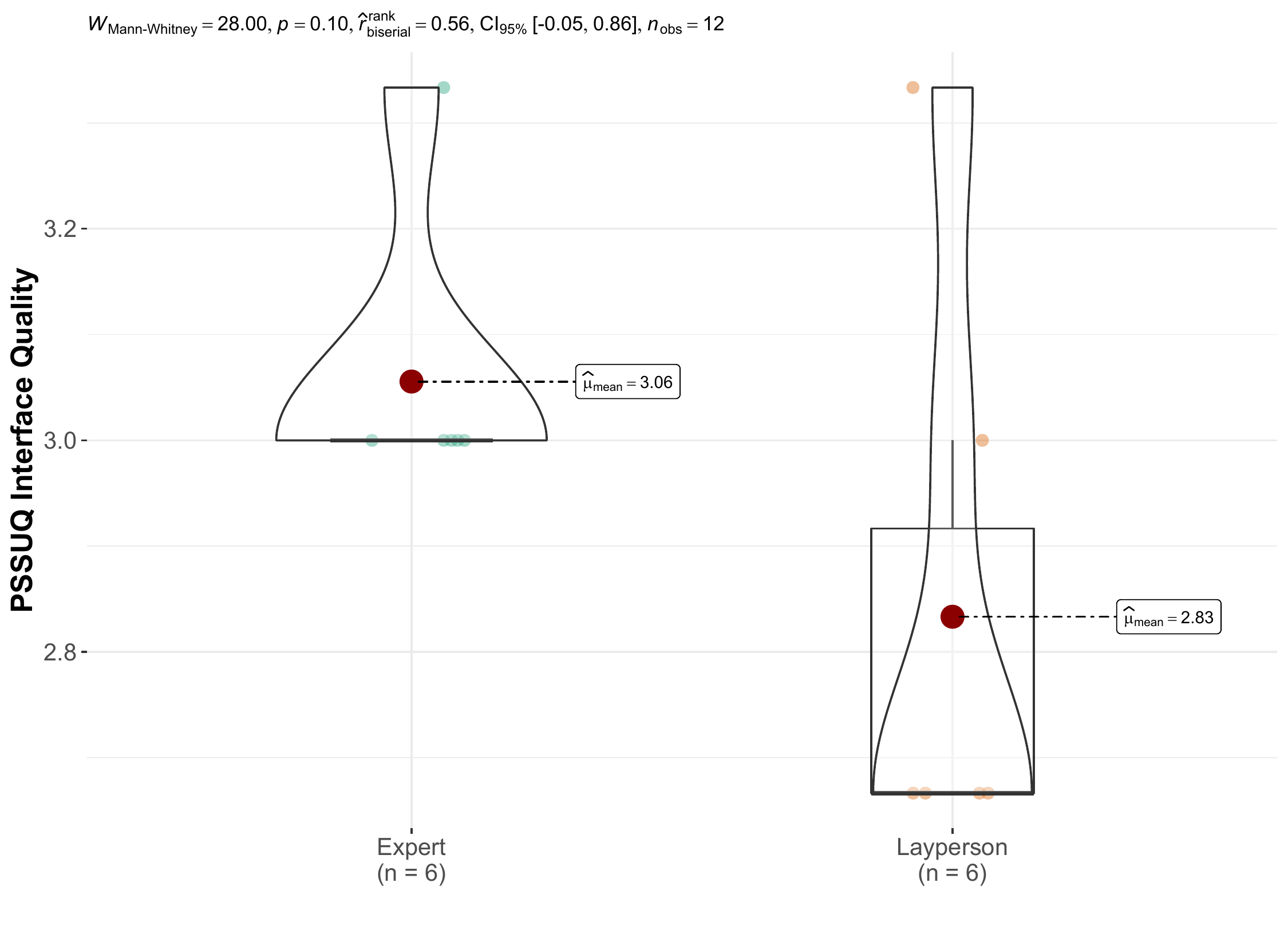}
        \subcaption{SAM Interface Quality}
        \label{fig:pssuq_interqual}
    \end{subfigure}
    \begin{subfigure}[b]{0.45\textwidth}
        \centering
        \includegraphics[width=\textwidth]{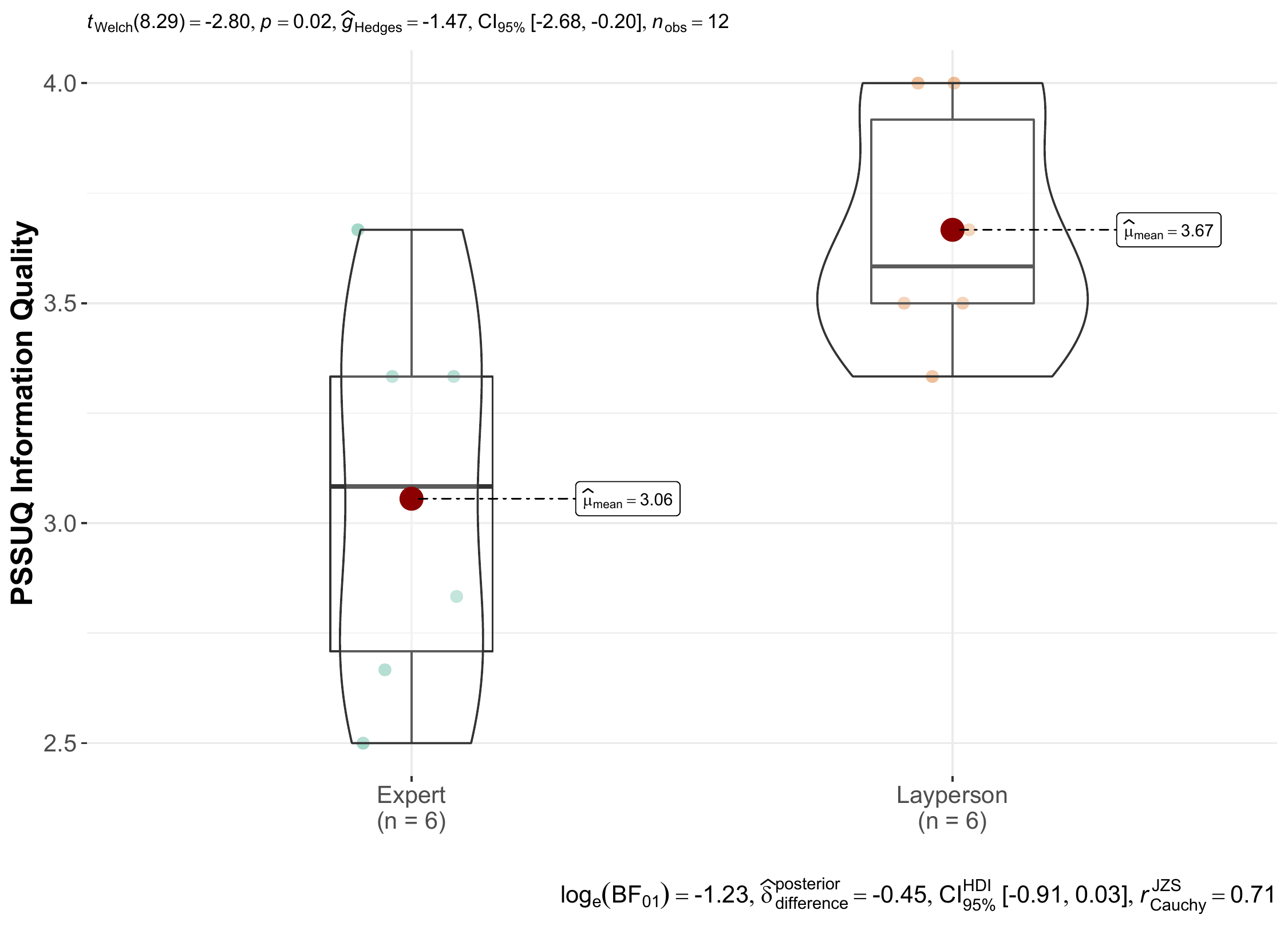}
        \subcaption{SAM Information Quality}
        \label{fig:pssuq_infoqual}
    \end{subfigure}
        \begin{subfigure}[b]{0.45\textwidth}
        \centering
        \includegraphics[width=\textwidth]{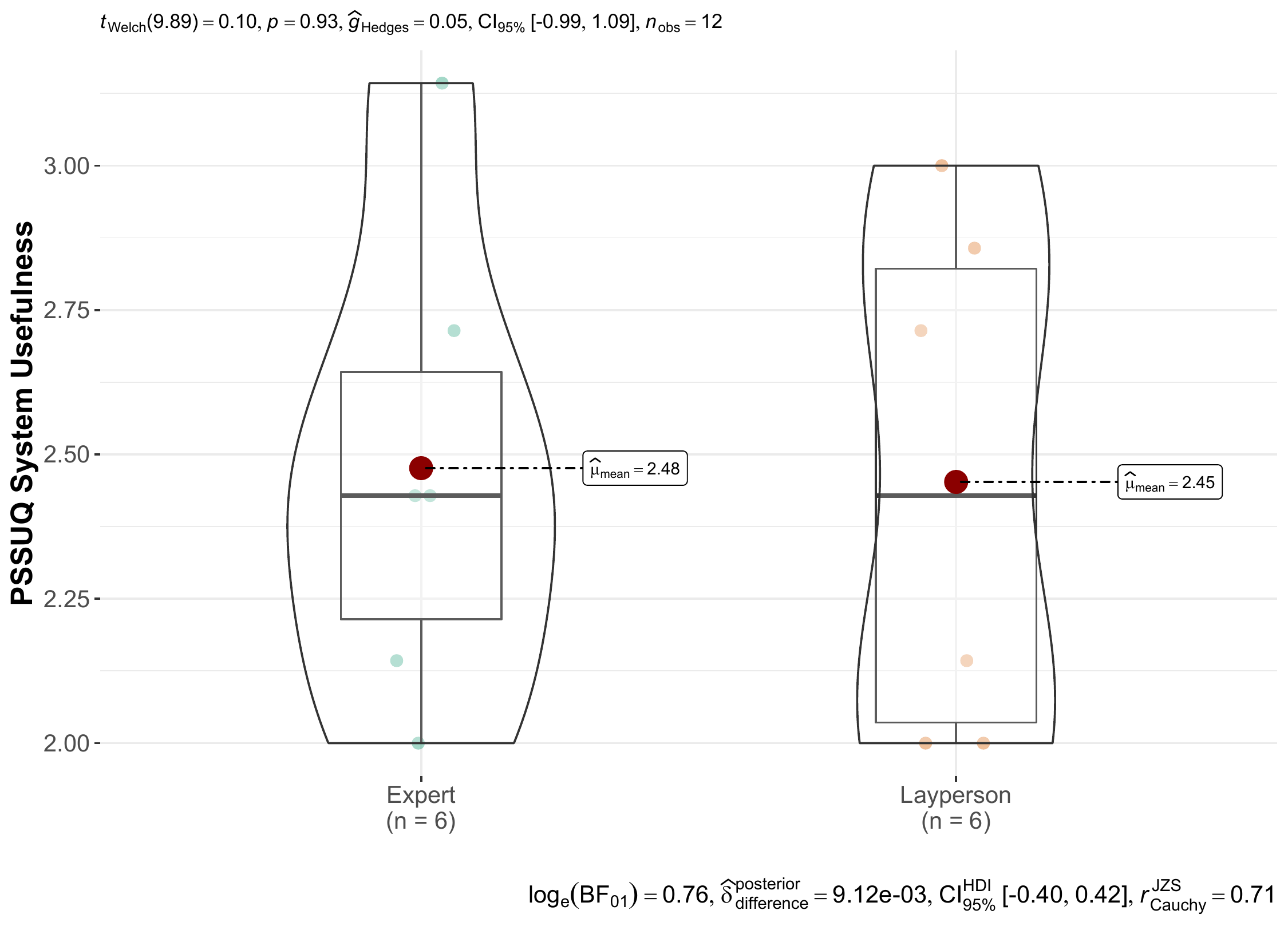}
        \subcaption{SAM System Usefulness}
        \label{fig:pssuq_sysuse}
    \end{subfigure}
    \caption{Comparison between layperson and experts for the PSSUQ}
    \label{fig:pssuq_scores}
\end{figure}

%\paragraph{PSSUQ:}
%Figure~\ref{fig:pssuq} shows that experts and laypersons rated the application as usable. 
Sauro and Lewis~\cite{sauro2016quantifying} report an overall mean of 2.82 in 21 usability studies. The overall mean of our application is \m{2.89} (\sd{0.25}), indicating a usable application (see~\autoref{fig:pssuq_scores}).
We found significant differences between experts and layperson for the subscale \textit{information quality} (see~\autoref{fig:pssuq_infoqual}). Experts rated the quality of the information significantly better.

%SYSUSE: 2.80
%INFOQUAL: 3.02
%INTERQUAL: 2.49
%Overall: 2.82

%Sauro and Lewis (2016) involving 21 studies and 210 participants may be helpful in interpreting the scores you get from your research:
%SYSUSE: 2.80 INFOQUAL: 3.02 INTERQUAL: 2.49 Overall: 2.82

\paragraph{Open Feedback:}
Participants highlighted the simplicity: \say{intuitive} [P2], %\say{easy operation} [P3],
\say{easy to use} [P8]. The concept \textit{social navigation} was appreciated: \say{the idea is very convenient because it's very interesting to see the position of the other drivers} [P10].
No major functionality was missed. % ([P2]: \say{everything else is nice to have}). 
Additional information such as distance to other drivers [P9] and integration in Google Maps [P11]. % and a button for aborting navigation ([P4], [P12]) was requested.

\section{Discussion}
Participants agreed that a supporting application can improve the sense of belongingness while addressing issues of collaborative driving.

\subsection{Target Automation Level}
%The focus group was conducted without referring to a special automation level~\cite{sae2014taxonomy}. 
The prototype is targeted towards lower automation levels, however, the concept is transferable to high automation levels as \textit{location sharing} and \textit{media sharing and entertainment} will still be relevant.
Navigation in AVs is likely to have lower importance as simple input of the target might suffice. Knowing where the other participants of a journey are, however, will still be relevant. While we used a smartphone application, this functionality could easily be integrated in today's large screens (e.g., MBUX Hyperscreen~\cite{mbux2021hyperscreen}).

\begin{comment}

    \begin{figure}[ht!]
        \centering
        \includegraphics[width=0.50\textwidth]{figures/navigation_vs_social.jpg}
        \caption{Hypothesized growing importance of fulfilled social needs in AVs based on SAE's definition of level of automation~\cite{sae2014taxonomy}}
        \label{fig:nav_vs_social}
    \end{figure}
    
\end{comment}

\subsection{Future Activities During Automated Driving}
We agree that activities will fundamentally change in higher automation levels~\cite{pfleging2016investigating}. While it unclear how future passengers will actually behave, we believe \textit{social needs} will become more important. Today, airline passengers already experience a fully autonomous journey. It was hypothesized that the flying experience could be influenced by the interaction with others, however, there seems to be \say{little structured research}~\cite[p. 14]{patel2018passenger}. Our focus group revealed an interesting pattern in passenger behavior, namely starting movies synchronized, that might be transferred to AVs.

\subsection{Limitations}
The number of participants in the study was of moderate~\cite{10.1145/2858036.2858498} size (\N{12}). Also, the number of the focus group participants was low (\N{5}). Therefore, findings might not be generalizable. However, we believe the topics that emerged to be of greater relevance. The application prototype currently does not support \textit{media sharing and entertainment}. The preliminary evaluation of the application did not take place in a vehicle. Also, the ``friend'' that was interacted with was realized via Wizard-of-Oz. Nevertheless, ratings regarding usability should still be valid. However, the usage has to evaluated during a ride. 

\section{Conclusion \& Future Work}
Our work shows a novel direction for collaborative driving. We report on an expert focus group (\N{5}) revealing insights into social needs in driving such as location-awareness of fellow travelers. We suggested an application to aid in fulfilling these needs.
Results of a preliminary user study (\N{12}) showed that our approach is feasible and well-received.
In the future, we will enhance this prototype towards an integrated system serving multiple use cases.

\begin{acks}
The authors thank the company Elektrobit and all study and focus group participants.
This work was conducted within the project 'Interaction between automated vehicles and vulnerable road users' (Intuitiver) funded by the Ministry of Science, Research and Arts of the State of Baden-Württemberg.
\end{acks}

\bibliographystyle{ACM-Reference-Format}
\bibliography{sample}

\end{document}